# A SHORT-DISTANCE INTEGRAL-BALANCE SOLUTION
# TO A STRONG SUBDIFFUSION EQUATION:
# A Weak Power-Law Profile

### Jordan Hristov

**Abstract** – The work presents an integral solution of the time-fractional subdiffusion through a preliminary defined profile with unknown coefficients and the concept of penetration layer well known from the heat diffusion The profile satisfies the boundary conditions imposed at the boundary of the boundary layer in a weak form that allows its coefficients to be expressed through the boundary layer depth as unique parameter describing the profile. The technique is demonstrated by a solution of a time fractional subdiffusion equation in rectilinear 1-D conditions. *Copyright © 2009 Praise Worthy Prize S.r.l. - All rights reserved.*

## I. Introduction

The subdiffusion process has been observed in many real physical systems such as highly ramified media in porous systems [1],[2],[3],[4], diffusion in thick membranes [5], anomalous drug absorption and disposition processes [6].The underlying physics of subdiffusion is associated with a medium in which the mean square displacement of Brownian motion evolves on a slower-than-normal time scale, that is $\langle x^2 \rangle \sim t^\mu$, where the anomalous diffusion parameter is $\mu$ is within the range $0 < \mu < 1$. A continuum model of a subdiffusive material is consistent with the scenario in which the pore size is small in comparison with $\langle x^2 \rangle^{1/2}$ [1].

There has been a growing interest to investigate the solutions of subdiffusive equations and their properties for various reasons which include modeling of anomalous diffusive and subdiffusive systems, description of fractional random walk, unification of diffusion and wave propagation phenomenon, and simplification of the results. The common methods for solving fractional-order equations are purely mathematical, even tough they are approximate in nature, among them: in terms of Mittag–Leffler function [7], similarity solutions [8], [9], Green's function [10], operational calculus [11] and variational iteration method [11, 12].-

The present work refers to an integral solution commonly known as heat-balance integral [13], [14]. The core of the model is the assumption of the thermal penetration layer propagating with a finite velocity. Beyond the front of this layer the medium is undisturbed. This idea of Goodman [13], in fact, corrects the physical incorrectness of the parabolic heat-equation where the speed of the flux is infinite. The integral solution suggests a prescribed profile with unknown coefficients satisfying the boundary conditions at the both ends of the penetration layer. The integral approach to the fractional equation suggests replacement of the real function by an approximate profile and integration over the penetration depth. The technique was demonstrated recently [15] in a solution of a hall-time fractional equation resulting by splitting of the normal (diffusion) parabolic equation and Riemann-Liouville time-derivative.

The specific case reported in this article is an example demonstrating the technique of the integral method to which the method was already applied by a general parabolic profile [15].

## II. THE INTEGRAL METHOD

### II.1. Mathematical formulation

It is assumed that the temperature (concentration) $C(x,t)$ in the semi-infinite subdiffusive material satisfies the one-dimensional fractional diffusion equation given by

$$\frac{\partial^\mu C(x,t)}{\partial t^\mu} = D_\mu \frac{\partial^2 C(x,t)}{\partial x^2}, \ 0 < \mu < 1 \qquad (1)$$

$$C(0,t) = C_s, \ t \geq 0; \ C(x,0) = C_\infty, \ x > 0;$$

$$\frac{\partial C(x,0)}{\partial t} = 0, \ x > 0 \qquad (2a, b, c)$$

$$\frac{\partial^\mu C}{\partial t^\mu} = {}_{RL}D_t^\mu C(x,t) = \frac{1}{\Gamma(1-\mu)}\frac{d}{dt}\int_0^t \frac{C(x,\tau)}{(t-\tau)^\mu}d\tau \quad (3)$$

where $D_\mu$ is a sort of fractional diffusion coefficient of dimensions $[D_\mu] = [m^2/\text{sec}^\mu]$ and (3) is the Riemann-Liouville fractional derivative of $C(x,t)$ with respect to the time $t$ [10].

The common approach is to use fractional derivatives in the Caputo sense and Laplace transform, thus avoiding the problem to define a boundary condition containing the limit value of the Riemann-Liouville fractional derivatives at $t=0$. The integral method avoids this problem by a preliminary definition of the function approximating the distribution generated by the subdiffusion equation (1) across a limited in length penetration layer $\delta$, which permits the Riemann-Liouville definition of the fractional calculus to be used. The penetration layer evolves in time, i.e. $\delta(t)$ and beyond the point $x = \delta(t)$, the medium is undisturbed [13], [14], namely

$$C(x,\delta) = C_\infty, \; x > \delta \qquad (4a)$$

$$\frac{\partial C(\delta,t)}{\partial x} = 0 \; ; \; \delta(t) = 0, \; t = 0 \qquad (4b)$$

This approach is supported be experimental facts of almost sharp fronts of penetration of the diffusion substances [5], [6], [16], [17]. Moreover, the fractional diffusion equation referring sub-diffusion problems [5] the mass propagation (diffusion) is so slow [16], [17] that the concept of the penetration layer becomes essential. At any time $t$ the integral of both sides (1) along $\delta$ yields

$$\int_0^\delta \left[\frac{\partial^\mu}{\partial t^\mu}C(x,t)\right]dx = D_\mu \frac{\partial C}{\partial x}\bigg|_{x=0}^\delta \qquad (5)$$

The integral of left-side in (5) is termed **fractional-time Heat-balance Integral** (FT-HBI) following [15]. If the distribution $C(x,t)$ across the penetration layer is approximated by $C_a(x)$ depending only on $x$, $0 < x < \delta(t)$ and the boundary conditions (4a, b) and (2b) are applied, then we get a profile (distribution) expressed as a function of $x$ and coefficients depending on $\delta(t)$. Further, replacing $C(x,t)$ by $C_a(x,\delta)$, in (3b) we get

$$\int_0^\delta \left[\frac{\partial^\mu}{\partial t^\mu}C_a(x,\delta(t))\right]dx = D_\mu \frac{\partial C_a(x,\delta(t))_a}{\partial x}\bigg|_{x=0}^\delta \qquad (6)$$

The evaluation of (5) through particular expression of $C_a(x,\delta)$ needs evaluation of the fractional-time derivative, namely

$${}_{RL}D_t^\mu C_a(x,\delta(t)) = \frac{1}{\Gamma(1-\mu)}\frac{d}{dt}\int_0^t \frac{C_a(x,\delta(t),\tau)}{(t-\tau)^\mu}d\tau \quad (7)$$

### II.2. Approximate profile

#### II. 2.1. Complete approximate profiles

The commonly used functions are polynomials of integer order: quadratic or cubic [13], [14] or a generalized parabolic profile $C_a(x,t) = a_1 + a_2(1+a_3x)^n$ with unspecified exponent [18], [19]. With $C(0,t) = C_s$ as boundary condition at $x = 0$ the profile simply results in

$$C_a(x,t) = C_\infty + (C_s - C_\infty)\left(1-\frac{x}{\delta}\right)^n \qquad (8a)$$

$$\Theta_a(x,t) = \frac{C-C_\infty}{C_s-C_\infty} = \left(1-\frac{x}{\delta}\right)^n \qquad (8b)$$

The profile satisfies the conditions of the penetration layer at any value of the exponent $n$ [15], [18], [19] which has to be determined through additional conditions [19], [20], [21].

#### II.2.2 Weak approximate profile and its characteristics

The weak approximate profile, employed in this work is a power-law relationship

$$\Theta_a(x,t) = \frac{C-C_\infty}{C_s-C_\infty} = 1-\left(\frac{x}{\delta}\right)^n \qquad (9)$$

The profile (9) satisfies the conditions $\Theta_a(0,t) = 1$ and $\Theta_a(\delta,t) = 0$ at any value of the exponent $n$. Moreover, $\partial\frac{\Theta_a(x,t)}{\partial x} = -n\frac{x^{n-1}}{\delta^n}$ that leads to $\partial\frac{\Theta_a(\delta,t)}{\partial x} = -\frac{n}{\delta}$ and $\partial\frac{\Theta_a(0,t)}{\partial x} = 0$. Hence, this profile does not satisfy the classical boundary conditions [13], [14]: a) $\partial\frac{\Theta_a(\delta,t)}{\partial x} = 0$ (see 4b) and $\partial\frac{\Theta_a(0,t)}{\partial x} \neq 0$. Because of that is it termed here **weak approximate profile (WAP)**.

The condition (4b) is basic one in the Goodman method. However, in the anomalous diffusion [5], the experiments do not show sharp fronts defined by (4b).

The experiments of Kosztolowicz [5], for example, define a value of the concentration $C(0.05)$ as 5% of $C_s(x=0)$ as a measurable threshold and a corresponding front $\delta_{0.05}(t)$. Hence, the condition (4b) can be avoided under the assumption that the other two (i.e. 2a, b) are exactly obeyed. Actually, this approach reduces the number of the conditions imposed on the approximate profile. The choice of the profile was motivated by the recent article of Voller [23] to fractional Stefan problem where a linear profile with $n=1$ was chose.

### *II.2.3. Double integration approach*

The problem with the improper derivatives commented in (2.2.2) allow to apply again integration over the penetration layer of both sides of (5), namely

$$\int_0^{\delta}\left\{\int_0^{\delta}\left[\frac{\partial^{\mu}}{\partial t^{\mu}}C(x,t)\right]dx\right\}dx=\int_0^{\delta}\left[\int_0^{\delta}D_{\mu}\frac{\partial^2 C(x,t)_a}{\partial x^2}\bigg|\,dx\right]dx$$

$$\int_0^{\delta}\left\{\int_0^{\delta}\left[\frac{\partial^{\mu}}{\partial t^{\mu}}C(x,t)\right]dx\right\}dx=D_{\mu}C(0,t) \qquad \text{(10a,b)}$$

## III. Solutions with WAP

### III.1. Fractional-time integral approximation

The approximation of the Riemann-Liouville fractional derivative through WAP is the first step, namely

$$_{RL}D_t^{\mu}\Theta_a(x,t)=\,_{RL}D_t^{\mu}\left[1-x^n\delta^{-n}\right]=$$

$$=\frac{1}{\Gamma(1-\mu)}\frac{d}{dt}\int_0^t\left[1-\left(\frac{x}{\delta}\right)^n\right]\frac{1}{(t-\tau)^{\mu}}d\tau \qquad \text{(11a)}$$

Denoting the integral from 0 to $t$ in (11a) as

$$\int_0^t\frac{\left(1-x^n\delta^{-n}\right)}{(t-\tau)^{\mu}}d\tau=\Phi(t), \qquad \Phi(0)=\int_0^t\frac{1}{(t-\tau)^{\mu}}d\tau,$$

$$\Phi(\delta)=0 \qquad \text{(11b)}$$

and applying the Leibniz rule for differentiation under the integral sign, we get

$$\int_0^{\delta}D_t^{\mu}\Theta_a(x,t)=\frac{1}{\Gamma(1-\mu)}\int_0^{\delta}\frac{d}{dt}\Phi(t)dx=$$

$$=\frac{1}{\Gamma(1-\mu)}\left[\frac{d}{dt}\int_0^{\delta}\Phi(t)dx-\Phi(\delta)\frac{d\delta}{dt}\right] \qquad \text{(11c)}$$

With $\Phi(\delta)=0$, we have

$$\int_0^{\delta}D_t^{\mu}C_a(x,t)=\frac{1}{\Gamma(1-\mu)}\frac{d}{dt}\int_0^{\delta}\Phi(t)dx=$$

$$=\frac{1}{\Gamma(1-\mu)}\frac{d}{dt}\left[\left(\delta-\frac{\delta^{n+1}}{n+1}\right)\int_0^t\frac{1}{(t-\tau)^{\mu}}d\tau\right] \qquad \text{(11d)}$$

The expression (11d) yields

$$\frac{1}{\Gamma(1-\mu)}\frac{d}{dt}\left[\left(\delta-\frac{\delta^{n+1}}{n+1}\frac{1}{\delta^n}\right)\int_0^t\frac{1}{(t-\tau)^{\mu}}d\tau\right]= \qquad \text{(12a)}$$

$$=\frac{1}{(1-\mu)\Gamma(1-\mu)}\frac{d}{dt}\left[\left(\frac{n\delta}{n+1}\right)t^{1-\mu}\right] \qquad \text{(12b)}$$

Hence, the first integration of the LHS of (10a) from 0 to $\delta$ yields

$$\int_0^{\delta}\left[\frac{\partial^{\mu}}{\partial t^{\mu}}\Theta_a(x,t)\right]dx=\frac{1}{\Gamma(2-\mu)}\frac{n}{n+1}\frac{d}{dt}\left(\delta t^{1-\mu}\right) \qquad \text{(13a)}$$

The second integration means

$$\int_0^{\delta}\int_0^{\delta}\left[\frac{\partial^{\mu}}{\partial t^{\mu}}\Theta_a(x,t)\right]dx\,dx=\int_0^{\delta}\left[\frac{1}{\Gamma(2-\mu)}\frac{n}{n+1}\frac{d}{dt}\left(\delta t^{1-\mu}\right)\right]dx \qquad \text{(13b)}$$

That yields

$$\int_0^{\delta}\int_0^{\delta}\left[\frac{\partial^{\mu}}{\partial t^{\mu}}\Theta_a(x,t)\right]dx\,dx=\frac{1}{\Gamma(2-\mu)}\frac{n}{n+1}\frac{\delta d}{dt}\left(\delta t^{1-\mu}\right) \qquad \text{(13c)}$$

### III.2. Penetration depth

Further, taking into account the Right-Hand Side (RHS) of (10b) expressed through the approximate profile, we get

$$\frac{1}{\Gamma(2-\mu)}\frac{n}{n+1}\frac{\delta d}{dt}\left(\delta t^{1-\mu}\right)=D_{\mu} \qquad \text{(14a)}$$

From (13a) we have an equation governing the time evolution of the front of the penetration layer

$$\frac{\delta d}{dt}\left(\delta t^{1-\mu}\right)=ND_{\mu} \qquad \text{(14b)}$$

$$N=(n+1)\Gamma(2-\mu)/n \qquad \text{(14c)}$$

With the transform $\delta t^{1-\mu} = Z(t)$ [22] the equation (14b) reads

$$\frac{d}{dt}Z^2 - 2ND_\mu t^{1-\mu} = 0 \qquad (15a)$$

$$Z^2 = 2ND_\mu t^{2-\mu} + P_1 \Rightarrow \delta^2 = 2ND_\mu t^\mu + \frac{P_1}{t^{2-2\mu}} \qquad (15b)$$

that leads to the general solution

$$\delta = \sqrt{D_\mu t^\mu}\sqrt{\frac{2N}{2-\mu} + \frac{P}{t^{1-\mu}}} \qquad (15c)$$

The initial condition $\delta(t=0)=0$ leads to $P_1 = 0$. This is a physically defined requirement relevant to the fact that the source providing the mass at $x=0$ is of a finite power and the mass penetrates slowly into the medium. Therefore, the penetration layer depth can be expressed as

$$\delta = \sqrt{D_\mu t^\mu}\sqrt{\frac{2N}{2-\mu}} = \sqrt{D_\mu t^\mu}\sqrt{F_n(n)j_\mu} \qquad (16a)$$

$$F_n(n) = 2\frac{(n+1)}{n}, \quad j_\mu = \frac{\Gamma(2-\mu)}{(2-\mu)} \qquad (16b, c)$$

### III. 3.   Approximate profile
With the result (15a) we get

$$\Theta_a(x,t) = 1 - \left(\frac{x}{\sqrt{D_\mu t^\mu}\sqrt{F_n j_\mu}}\right)^n \qquad (17)$$

Through the similarity variable $\eta = x/\sqrt{D_\mu t^\mu}$, we have

$$\Theta_a(x,t) = 1 - \left(\frac{\eta}{\sqrt{F_n j_\mu}}\right)^n \qquad (18)$$

The profile (17) defines the so-called *penetration stage*. If the medium is finite, a slab for example, of a length $L$, then the condition $L = \sqrt{D_\mu t^\mu}\sqrt{F_n(n)j_\mu}$ defines the critical time

$$t_L = \left(\frac{L^2}{D_\mu}\right)\frac{1}{F_n(n)j_\mu}. \qquad (19a)$$

Beyond $t \geq t_L$ the profile becomes simpler

$$\Theta_a^L(x,t) = 1 - (x/L)^n. \qquad (19b)$$

This stage is beyond the scope of the present work, so let us go back to the case when $t < t_L$.

### III.5.    Optimal exponent
### III. 5.1. Mean-squared error approach
The approximate should satisfy the domain equation, so the mean-squared error of approximation should be minimal, namely

$$E_\mu(t) \equiv \int_0^{\delta(t)}\left[\frac{\partial^\mu}{\partial t^\mu}\Theta_a(x,t) - D_\mu\frac{\partial^2\Theta_a}{\partial^2 x}\right]^2 dx \geq 0,$$

$$E_\mu(t) \to \min \qquad (20a)$$

and

$$e_\mu(x,t) = \frac{\partial^\mu}{\partial t^\mu}C_a(x,t) - D_\mu\frac{\partial^2 C_a}{\partial^2 x}$$

$$E_\mu(t) \equiv \int_0^{\delta(t)}\left[e_\mu(x,t)\right]^2 dx \qquad (20b,c)$$

With the dimensionless profile (16a) we have

$$\frac{\partial^\mu}{\partial t^\mu}\Theta_a(x,t) = \frac{1}{\Gamma(1-\mu)}\frac{d}{dt}\int_0^t\left[1 - \left(\frac{x}{\delta}\right)^n\right]\frac{1}{(t-\tau)^\mu}d\tau =$$

$$= \left[1 - \left(\frac{x}{\delta}\right)^n\right]\frac{1}{(1-\mu)\Gamma(1-\mu)}t^{-\mu} \qquad (21a)$$

$$\frac{\partial^2\Theta_a}{\partial x^2} = -\frac{n(n-1)}{\delta^n}x^{n-2} \qquad (21b)$$

Then,

$$e_\mu(x,t) = \left[1 - (x/\delta)^n\right]\frac{1}{\Gamma(2-\mu)}t^{-\mu} + \frac{n(n-1)}{\delta^n}x^{n-2} \qquad (22)$$

$$E_\mu(t) = \left[\frac{1}{\Gamma(2-\mu)}\right]^2 t^{-\frac{3}{2}\mu} \times$$

$$\times\left[\frac{(2n+1)(n+1) + (n+1) - 2(2n+1)}{(2n+1)(n+1)}\right] \times$$

$$\times\sqrt{D_\mu F_n j_\mu} +$$

$$+ n^2(n-1)^2\left(\frac{1}{2n-3}\right)\frac{1}{\delta^3} +$$

$$2\frac{n}{\Gamma(2-\mu)}t^{-\mu}\delta -$$

$$-2\frac{1}{\Gamma(2-\mu)}(n-1)\frac{1}{\left(\sqrt{D_\mu F_n j_\mu}\right)^n t^{n\frac{\mu}{2}+1}} \qquad (23)$$

All the terns of $E_\mu(t)$ are time-dependent and decreases in time; however it is impossible to detect which of them is the controlling one. If the error function $E_\mu(t)$ will be expressed as $E_\mu(t) \sim \dfrac{1}{\delta^3}$, that is $E_\mu(t) \sim t^{-\frac{3}{2}\mu}$ as in the classical method to minimize the error of approximation with $\mu = 1$ [20],[21], the only term becoming time-independent is the 2$^{nd}$ one in (23). However, it goes to zero at $n = 1$, that leads to a linear profile, that in, fact is unacceptable. Therefore, the direct application of the mean-squared error approach to the case concerning the weak profile approximation of the solution of the fractional-time anomalous diffusion equation failed. In contrast, when the complete profile (8) approximates the solution [22], then the approach provides that the exponent depends on the fractional order, i.e. $n = n(\mu)$. Obviously, the choice of the approximating profile is crucial in the heat-balance integral solution of fractional subdiffusion equations.

### III. 5.2.  Some limits of $n = n(t, \mu)$

Following the idea mentioned in the previous point let us express (23) in the form $E_\mu(t) = \left(1/\delta^3\right) K_1\left(n, t, \mu\right)$, that is $E_\mu(t) \equiv \left(1/t^{(3/3)\mu}\right) K_1\left(n, t, \mu\right)$ we get

$$K_1(n,t,\mu) = \begin{cases} \left[\dfrac{1}{\Gamma(2-\mu)}\right]^2 \times \\[2mm] \times\left[\dfrac{(2n+1)(n+1)+(n+1)-2(2n+1)}{(2n+1)(n+1)}\right]\times \\[2mm] \times\sqrt{D_\mu F_n j_\mu} + \\[2mm] +n^2(n-1)^2\left(\dfrac{1}{2n-3}\right) + \\[2mm] +2\dfrac{n}{\Gamma(2-\mu)}D_\mu F_n j_\mu - \\[2mm] -2\dfrac{1}{\Gamma(2-\mu)}(n-1)\dfrac{1}{\left(\sqrt{D_\mu F_n j_\mu}\right)^n t^{n\frac{\mu}{2}+1}}\delta^3 \\[2mm] + \end{cases}$$

$$(24)$$

In $K_1(n, t, \mu)$ only the last term is proportional to $t^{(3/2)\mu}$, so the minimization with respect to $n$ exclude this term. With this simplification and varying the fractional order from $0.1$ to $0.9$ we get values of the exponent $n_\mu$ (Performed by Maple 13) summarized in Table 1.

For seek of simplicity calculations, a numerical experiment was performed with $D_\mu = 1$ which enable to perform the first approximation about the functional relationship $n(\mu)$, even though this is by far a way from the real values of $D_\mu$. However, the numerical results performed further in this work show that the exponent is in the range $1 \le n_\mu \le 1.5$.

| | | Optimal Exponent $n_\mu$ [-]* | | |
|---|---|---|---|---|
| Fractional order $\mu$ [-] | Fractional correction factor $j_m$ [-] | $D_\mu = 1.10^{-9}$ ($m^2/s^\mu$) **A** | $D_\mu = 6.3.10^{-10}$ ($m^2/s^\mu$) **B** | $D_\mu = 1$ **C** |
| 0.1 | 0.506 | 1 | 1.006 | 1.464 |
| 0.2 | 0.517 | 1.007 | 1.006 | 1.466 |
| 0.3 | 0.534 | 1.007 | 1.006 | 1.468 |
| 0.4 | 0.558 | 1.007 | 1.006 | 1.469 |
| 0.5 | 0.590 | 1.007 | 1.006 | 1.471 |
| 0.6 | 0.633 | 1.007 | 1.007 | 1.472 |
| 0.7 | 0.690 | 1.007 | 1.007 | 1.474 |
| 0.8 | 0.765 | 1.007 | 1.007 | 1.475 |
| 0.9 | 0.864 | 1.007 | 1.007 | 1.477 |
| Note: The values of $D_\mu$ are taken from [24],[25],[26] for thick membranes at $\mu = 0.9$ : **A**-glucose ; **B**-sucrose ; **C**-(any dimension) | | | | |

The results in Table 1 reveal that irrespective of the variations in $\mu$ the minimization procedure provides, in fact, a linear profile with $n_\mu \approx 1$. This is reasonable results since with $D_\mu \sim 10^{-10}$ the 1st and the 3$^{rd}$ terms in $K_1(n, t, \mu)$ are practically negligible. Then, from the only remaining term we get

$$n^2(n-1)^2\left(\frac{1}{2n-3}\right) = 0 \Rightarrow n = 1 \qquad (25a)$$

In this context, let us look again at the expression of $E_\mu(t)$ and take into account the order of magnitudes of all terms. With $\delta \equiv D_\mu^{1/2} \sim 10^{-5}$, the first term is growing in time and the minimization with respect to $n$ provides

$$(2n+1)(n+1)+(n+1)-2(2n+1) = 0 \quad (25b)$$

resulting in $n = 0$ which is unrealistic. With $n = 1$ we get for (25b) a value of $2$.

The second and the last terms of $E_\mu(t)$ goes to zero at $n = 1$. Therefore, with $n \approx 1$ we can attain an almost minimal value of the mean-squared error. As the time goes on, i.e. the contribution of the fist and the 3$^{rd}$ term of $E_\mu(t)$ become significant and we may suggest that in this case $n > 1$ and the last term (with a negative sign) compensates the others. The latter becomes evident from the numerical experiments performed (see next) where the optimal value of $n$ is within the range $1 \le n \le 1.5$ when $0.5 \le \eta \le 1$.

The choice $D_\mu = 1$ simplifies the calculations and allows getting adequate approximation for $0.3 < \mu < 0.5$ as it will be demonstrated by the numerical experiments performed.

### III. 5.3.    Numerical experiments with a fixed exponent

Some numerical tests valuating the above performed estimation of the exponent $n$ are shown in Fig.1. Two exact solutions at $\mu = 1/2$ and $\mu = 1/3$ were used

$$M_{1/2}(z) = \frac{1}{\sqrt{\pi}} \exp\left(-z^2/4\right) \qquad (26a)$$

$$M_{1/3}(z) = 3^{2/3} \, Ai\left(\frac{z}{3^{1/3}}\right) \qquad (26b)$$

Expressed through the auxiliary M-Wright function $M_\mu(z)$ [23]

$$M_\mu(z) \equiv \sum_{k=0}^{\infty} \frac{z^k}{k! \, \Gamma\left[-\mu k + (1-\mu)\right]} \qquad (27c)$$

$0 < \mu < 1$

where $Ai$ is the Airy function.

The plots reveal that the linear profile is an adequate approximation of the solutions to the time-fractional subdiffusion equations with fractional orders $0 < \mu \le 2$ (Fig.1a, b and Fig.2a,b) and to some extent to solutions with $\mu = 0.3 - 0.4$ (Fig.1c and Fig.2c). The increase in the fraction order of the subdiffusive system needs higher values of $n_\mu$ as it illustrated by the plots in Fig.1d and Fig.2d.

## IV. Some comments on the relevance of the developed solutions

The developed solutions show that the weak profile $\left[1 - (x/\delta)^n\right]$ can be applied to derive approximate solutions to the fractional-time anomalous subdiffusion equation. The solution can be developed for arbitrary value of the exponent $n$ because neither the concept of the penetration layer neither the heat-balance integral method imposes additional conditions.

The evaluation of the optimal exponent through the mean-squared error approach requires the approximate profile to satisfy the domain equation. This approach results in cumbersome expressions but some terms of the expressions can be avoided if real values of the fractional diffusion coefficient $D_\mu \sim O(10^{-9})$ are taken into account. With $D_\mu$ of this order of magnitude, the outcome is a linear profile irrespective of the values of the fractional parameter $\mu$. However, the numerical results indicate that this profile is adequate for cases $0.1 < \mu < 0.3$ with, while $n \approx 1.25 - 1.5$ allow s to get good approximations with $0.35 < \mu < 0.5$ within the range $0 < \eta < 0.5$. The increase in $\eta \to 1$ and $\mu \to 0.5$ requires the exponent to increase toward $n \approx 2$ and $n \approx 3$. The plots in Fig.1d and Fig.2d show very well these numerical estimations.

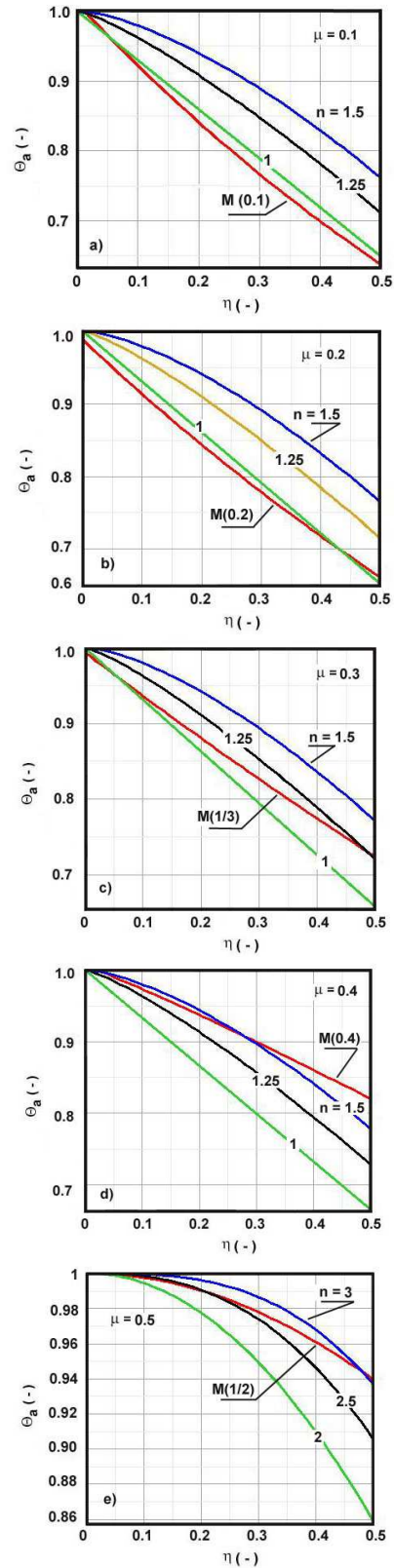

Fig. 1.  Approximate profiles at $0.1 < \mu < 0.3$ with fixed exponents of the profile and short distances expressed through the similarity variable $0 < \eta < 0.5$. The exact solution developed by the auxiliary M-function is denoted by $M(\mu)$.

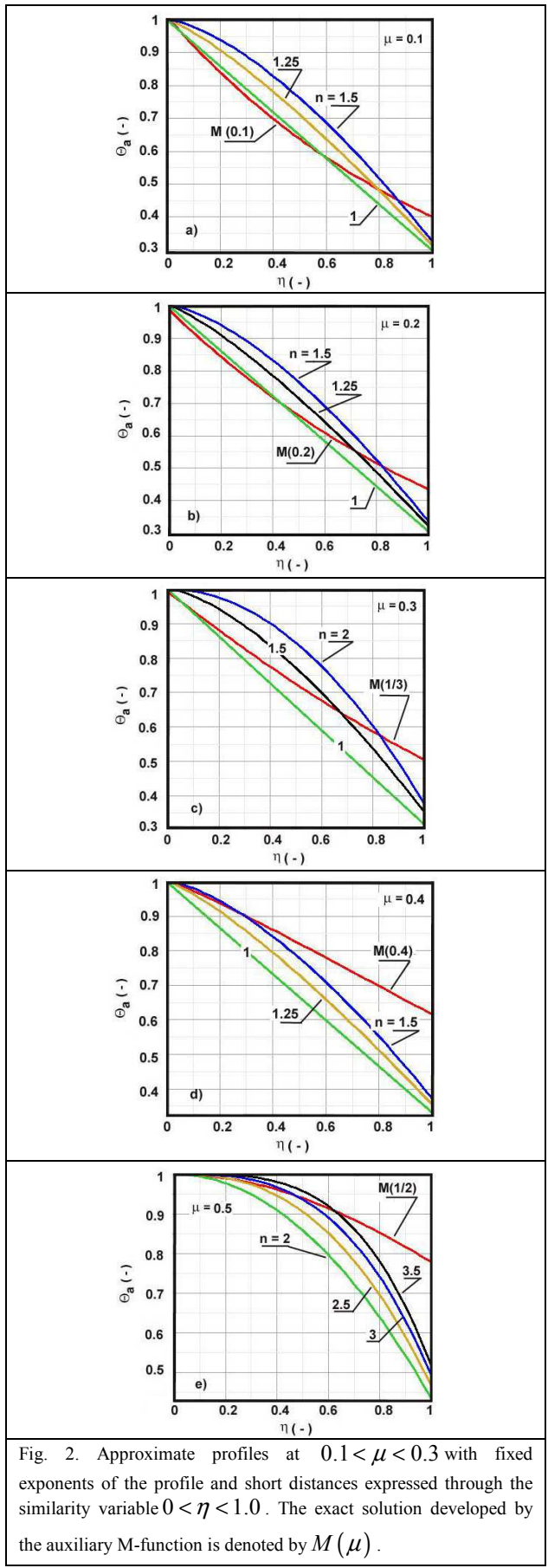

Fig. 2. Approximate profiles at $0.1 < \mu < 0.3$ with fixed exponents of the profile and short distances expressed through the similarity variable $0 < \eta < 1.0$. The exact solution developed by the auxiliary M-function is denoted by $M(\mu)$.

The numerical experiments were performed with small values of the similarity variable $\eta$ in to ranges: $0 \leq \eta \leq 0.5$ and $0 \leq \eta \leq 1.0$. The ranges were chosen as the best ones where the suggested approximate profile provides solutions closet to the exact ones. From physical point of view, with $D_\mu \sim 10^{-9} \left( m^2 / s^\mu \right)$, $0.5 < j_\mu < 0.9$ and $n_\mu \approx 1$ we get $\eta \approx 0.315 \times 10^4 \left( x t^{-\mu/2} \right)$. With $t = 1s$ and $\eta = 0.5$ we have $x \approx 1.587 m$, while for $x \approx 1mm$ (the case of thick membranes), we have a time about $t^{\mu/2} \approx 0.158 s^{\mu/2}$. The first case ($x \approx 1.587 m$) could be considered as a near-surface distance in large geological structures (porous media) described by the time-fractional subdiffusion equation. In the same context, the second limit $\eta = 1$ used in the numerical experiments, reveals: with $t = 1s$ we have $x \approx 3.174 m$, while for $x \approx 1mm$ we get $t^{\mu/2} \approx 3.174 s^{\mu/2}$.

The cases of thick membranes [24], [25], [26] provide a thickness of the near membranes layer of about $1mm$ at about $500 s$. With the developed solution and $\mu \approx 0.9$ we have $\eta \approx 1.41$. Similarly, a thickness of $3mm$ and $t \approx 1500 s$ results in $\eta \approx 2.44$. Hence, with an upper limit of $\eta \approx 0.5$ the approximate solution is adequate for the first 20% of the near membrane layer ($x = 0$ is at the membrane wall). However, we have to bear on mind that the numerical results just commented corresponds to $\mu \approx 0.9$ and reliable data about subdiffusive parameters for $0.1 < \mu < 0.5$ are still missing in the literature.

The error of approximation by the weak power-law profile is within the range $2.5 - 4.5\%$ for $0 < \eta < 0.5$.

## V. Conclusions

The developed solutions of the time-fractional subdiffusion equation through a weak approximate profile and the heat-balance integral method reveal some features, among them:

a) The derivation of the profile is simple through substitution of the real function by an approximate profile in the time-fractional derivation of the Riemann-Liouville sense. This allows defining an equation about the time evolution of the front of the penetration layer.

b) The solutions are valid for the near-surface domain under a strong subdiffusion, i.e. $\mu < 0.5$.

c) The optimal profile needs determination of the optimal exponent through an additional condition imposed on the mean-squared error of the domain equation. This method provides adequate values of the exponent for $0.1 < \mu < 0.3$ and within this range of the similarity variable $0 < \eta < 0.5$.

d) The increase in the fractional order beyond $\mu \approx 0.3$ towards $\mu \approx 0.5$ needs exponents ranging from $1.5$ to $3$.

e) The weak profile used in this work can be considered as a weak surrogate of the complete parabolic profile which also approaches the case of $n \approx 1$ when the subdiffusive conditions are stronger.

f) The error of approximation by the weak power-law profile is within the range $2.5 - 4.5\%$ for $0 < \eta < 0.5$.

### Acknowledgments

The work was supported by the grant N10751/2010 of Univ. Chemical Technology and Metallurgy (UCTM), Sofia, Bulgaria.

## Nomenclature

$C(x,t)$ -concentration (arbitrary units)

$D_\mu$ -fractional diffusivity coefficient, ( $m^2/s^\mu$ )

$E_\mu(t)$ - error function defined by eq.(19), (-)

$e_\mu(x,t)$ -function expression the domain equation through the approximate profile, eq.(20), (-)

$F_n$ -function defined by eq.(16b)

$I_{fr}$ -integral defined, (-)

$j_\mu$ -fractional correction factor, (-)

$n$ - unspecified exponent , (-)

$n_\mu$ - exponent of the approximate profile at a given fractional order $\mu$, (-)

$t$ - time , [s]

$C_s$ - surface concentration (boundary condition), (arbitrary units)

$C_\infty$ - temperature of undisturbed medium, (arbitrary units)

$x$ -space co-ordinate, [m]

$Z(t)$ - a variable defined by eq.(15)

*Greek letters*
$\Gamma$ -Gamma function, (-)

$\delta$ - depth of the penetration layer at $\mu = 1$, (m )

$\delta_\mu(t)$ - depth of the penetration layer at $0 < \mu < 1$, (m )

$\Phi(t)$ -function defiend by eq.11b , (-)

$\eta = x/\sqrt{\sqrt{D_\mu t^\mu}}$ -fractional similarity variable, (-)

$\mu$ -fractional parameter (order) [-]

$\tau$ -dummy variable in the RL definition of time-fractional derivative, (-)

$\Theta_a(x,t)$ -approximate dimensionless radial profile, [-]

*Subscripts*
$a$ - approximate

$fr(\mu)$ -fractional at any order of $0 < \mu < 1$

$\mu$ -fractional (at any $\mu$ )

$RL$ - Riemann-Liouville

$s$ -surface (at $x = 0$ )

$\infty$ -undisturbed medium

*Superscripts*
$a$ - approximate

# Authors' information

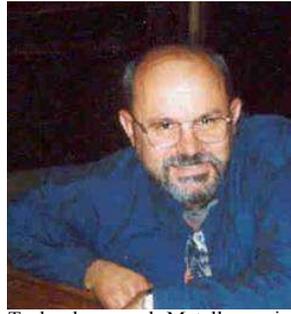

**Jordan Hristov** is associate professor of Chemical Engineering at the University of Chemical Technology and Metallurgy, Sofia, Bulgaria. He was graduated in 1979 as Electrical Engineer (MS equivalent) at the Technical University, Sofia, Bulgaria. His PhD thesis on the magnetically assisted fluidization was awarded by the University of Chemical Technology and Metallurgy in 1995. A/Prof. Hristov's research interests cover the areas of particulate solids mechanics, fluidisation, heat and mass transfer with special emphasis on scaling and approximate solutions. The main branch of his research is devoted to magnetic field effects of fluidization. Additionally, specific heat transfer topics are at issue, especially to thermal effects in accidents (fire). Relevant information is available at http://hristov.com/jordan .
.